# Embedded Spreadsheet Modelling

*Angela Collins*,
BDO Stoy Hayward LLP
3 Hardman Square, Manchester, M3 3EB
angela.collins@bdo.co.uk

**ABSTRACT**

*In larger accounting firms, specialist modellers typically sit in separate teams. This paper will look at the advantages of embedding a specialist modeller within a Corporate Finance Team.*

**1. Introduction**

This paper looks at the role of a modelling specialist based within the Corporate Finance (CF) Department of BDO Stoy Hayward LLP in Manchester. All other modellers also sit within Corporate Finance but are based together in London. It appears to be a feature of most of the larger accounting firms that modellers are based in separate teams rather than being distributed throughout the business. Modelling seems to be viewed in a similar way to Audit – not something which can be done alone. Due to this separation it can be difficult to improve the modelling skills of non specialist staff.

**2. Some issues**

**2.1 Image**

One of the issues which modellers face is the misconception that they are all technical specialists who do not understand the commercial realities under which most Corporate Financiers operate. There is the feeling that it is not possible to follow best practice due to time constraints. (An even more common issue is the lack of understanding of the importance of 'best practice'.)

**2.2 What is 'best practice'?**
It can be difficult to persuade non-specialists of the importance of 'best practice' especially as there is no definite agreement amongst modellers about what constitutes 'best practice'.

It seemed appropriate to focus on a small number of points at first so that non-modellers did not feel overwhelmed and therefore discouraged. As these points become embedded and their advantages seen then future points may be more easily accepted.

These basic points were - 1 – No circular references, 2 – No hard-coding of values within formula, 3 – Test as you go along, 4 – Formulae to be consistent. 5 – Separate inputs, calculations and outputs – make it clear which of these each cell is. 6 – Keep it simple!

**3. Corporate Finance Survey**

**3.1 Background**

A survey was conducted (via email) in BDO Corporate Finance in Manchester.

Participants were advised that a paper was being prepared which would look at the role of a modelling specialist embedded in CF for EuSpRIG and they were asked the following questions:
1) Have you found it useful having a modelling specialist based here?
2) Is there anything in particular that has helped?
3) Is there anything you would like done differently?
4) Have you changed the way you work with spreadsheets as a result of working with a modelling specialist?

**3.2 Responses**

Responses were received from 11 people from a team of 28.

**Q1 – Have you found it useful having a modelling specialist based here?**

All respondents agreed that having a modeller integrated into CF was useful.
The benefits include improved risk management, the ability to sell modelling as a stand-alone service, the ability to take on more complex modelling assignments and improved credibility with external institutions.

**Q2 – Is there anything in particular that has helped?**

The delivery of in-house training which has 'dramatically improved' Excel skills within the department. Availability of a named person as a trouble-shooter. Improved ability to resolve technical issues in-house. Educating CF about tools in Excel which have significantly cut down on time spent on financial analysis.

**Q3 – Is there anything you would like done differently?**

As there has been a growing understanding of the benefits that a specialist modeller offers there have been increasing requests for modelling support. This can be difficult to schedule when there is only one modeller in the office.

**Q4 – Have you changed the way you work with spreadsheets as the result of working with a modelling specialist?**

This included two main areas, 1 – Specific techniques (including how to use flags and how to used named ranges in calculations. 2 – Best Practice - this referred to in a number of responses. Particularly mentioned were improved version control and not hard-coding values into formulae.

**3.3 Isolation**

**Issues**

2Copyright © 2009 EuSpRIG & The Author(s)

One of the issues that must be addressed if you work on your own as a modeller is that of isolation. There are no immediate colleagues performing a similar role and this can be difficult, especially when there is a technical issue to resolve. It has been useful to build a strong relationship with other modellers within BDO. Frequent modelling updates are held and work is spread around the team where possible.

**4. Training**

**4.1 Formal**

There is an existing in-house modelling course which takes one day at the basic level but can run over two days, (this would include half a day examining how to review a model). It can be difficult for people to commit to two days, especially if they are concerned about their existing level of skills.

The course is currently being developed on a modular basis. This will enable us to tailor courses to a variety of audiences and it is hoped that this will increase attendance.

**4.2 Ad-hoc**

A-hoc training has been popular. Especially if staff have a problem when building a model. It is possible to fix the model and send it back and this would be the quickest solution at the time. However next time the issue arose the staff member would not be able to resolve it and the modeller would need to intervene again. It is more useful to sit with the person involved and show them one way of fixing the issue. Where possible alternative solutions that may be used in different circumstances are demonstrated. Training delivered in segments like this can be more easily absorbed and is seen to be directly relevant to the role that person performs (Mc Quire, 2007). There are some excellent external training courses available but a course is only the beginning. On-going support is crucial in developing modelling skills.

Although the general level of skills in the department has improved it is unrealistic to expect that everyone will develop into a modelling expert. The main aim is that people learn what not to do and when they need to ask for help.
Advanced Excel users can sometime forget what it is like to not know the basics. This became clear when a short course was presented to another department. Rather than the course content being dictated by what XXX the attendees were asked what they wanted to learn. Their responses included how to use shortcut keys, how to merge data from two columns into one and how to work with multiple sheets. Feedback after the course showed that there were two elements which were found to be particularly useful, one was general best practice and the other was the use of Edit – Replace. This is a function which most modellers will use without thinking but, as the course was running, it became clear that this was new to the attendees and it was therefore explained in detail. This flexibility and awareness has proved crucial. It is more effective to have simple techniques understood than to have advanced techniques only half absorbed.

**5. Assumptions/Model Review/Testing/Documentation**

When model are developed internally there is a reluctance to produce a specification, and those specifications which are produced are generally inadequate (Pryor 2003).

Many modellers have had the experience of building a model which implements all the assumptions supplied.  A sophisticated, well constructed and robust model is presented to the client.  They look at the results and believe that the model is wrong.  What they actually mean is they don't like the results.  It is therefore crucial that all conversations regarding a model are documented and that you can demonstrate to the client how their assumptions have been implemented.

Providing full documentation is perceived to be onerous and something which can be allowed to slip when under time-pressure.  Once simple way to increase ease of documentation is the use of cell comments.  A simple macro can be run which will extract all the cell comments from a model and this can form a basic documentation.

Users are encouraged to perform a substantial amount of testing. (Pryor, 2004 – 'Testing is the only way to tell what the spreadsheet actually does').  One technique which is popular is to use unrealistic inputs.  For example it is unlikely that fixed assets will depreciate over 2 months.  It is however a useful way of checking that your straight line depreciation works and doesn't over depreciate.  It is also worth deleting all your inputs (providing you have made a back-up copy of your model first)!  Then run the model with only single sections of inputs complete and check that the results are as expected.  One of the errors frequently found is that cashflows occur in the wrong period.  A quick way to test this is to use excessive inputs.  i.e. if there is a fixed asset purchase of £1,000 add a lot of noughts to the end.  When you look at the Financial Statements it will be clear if the timing is correct.

**6. Developments**

**6.1 Template models**

It is tempting to pick up a model which has been developed for a previous job and amend that.  This poses a number of risks: the model may only be appropriate in some circumstances.  If you pick up a model which someone else developed you may not understand how it works.  Also, if a model has been used previously there is a tendency to assume that it works and there is a reluctance to invest further time & cost in testing.

**Principles of template models**

There needs to be a consensus about the content of a template model.  They should not be developed by one modeller who works in isolation.  When a model is built for an external client a model specification will (should) be approved.  – the same principle applies to template models built for use within the business.  The basic template is controlled by an identified member of staff to ensure that any development is appropriate and documented.

It is key that the limitations of a template model are understood. It is unlikely to cover every circumstance that may arise.  When a template model is picked up for use on a new job one approach is to ensure that all inputs have been cleared.  This prevents existing inputs being used in error and not updated when appropriate.

In common with other firms in the sector BDO have a base model which is now used for all PFI jobs.  Substantial time was spent in developing the model to ensure that it would meet standard specifications as these are generally well defined (Croll 2005).  When the first draft version was available a small group of PFI specialists gathered together and worked through the model, looking at all inputs and examining how these were taken through the calculations.

As the model is used in a variety of jobs it is further refined.  As these refinements arise the PFI team consider if they are specific and should therefore be confined solely to the job on which they arise.  If they are more generic changes then the base template model is updated and, where appropriate, these changes are applied to any other current jobs.

### 6.2 Code library

There are a number of elements which are common to many models developed.  The modelling team have recently developed a code library which will contain adaptable examples of code for these common elements such as depreciation, working capital etc. It is believed that this modular approach will improve the accuracy and speed of model builds.(Paine 2007).

### 6.3 Modelling Road-shows

The approach discussed has proved useful in the Manchester office.  The core modelling team are keen to develop links between 'official' modellers and other offices.  A leaflet has been developed and is being distributed to all offices.   On one side of the sheet there are some hints and tips, shortcuts etc.  The other side describes the services offered by the modelling team and contact details.  Once this leaflet has been distributed to an office an informal modelling road-show  will be held which will cover some best practice principles and encourage engagement with the modelling team.

## 7. Conclusion

Modelling experts do not operate in isolation.  They work so that clients, both internal and external, can analyse and understand information and make appropriate decisions.  We deliver the best product when we understand the environment in which these clients operate